\documentclass[%
 reprint,
 amsmath,amssymb,
 aps,
]{revtex4-2}

\usepackage{xcolor}
\usepackage{graphicx}
\usepackage{dcolumn}
\usepackage{bm}

\begin{document}

\preprint{APS/123-QED}
\title{Characterisation of a triple-species ${}^\textbf{87}$Rb/ ${}^\textbf{85}$Rb/ ${}^\textbf{133}$Cs\\  magneto-optical trap towards cold-atom interferometry}


\author{Mal Landru} \email{mal.landru@onera.fr} \affiliation{DPHY, ONERA, Université Paris-Saclay, 91120 Palaiseau, France}%
\author{Amandine Lauret} \affiliation{DPHY, ONERA, Université Paris-Saclay, 91120 Palaiseau, France}
\author{Yannick Bidel} \affiliation{DPHY, ONERA, Université Paris-Saclay, 91120 Palaiseau, France}
\author{Alexandre Bresson} \affiliation{DPHY, ONERA, Université Paris-Saclay, 91120 Palaiseau, France}
\author{Sylvain Schwartz} \affiliation{DPHY, ONERA, Université Paris-Saclay, 91120 Palaiseau, France}
\author{Alexis Bonnin} \affiliation{DPHY, ONERA, Université Paris-Saclay, 91120 Palaiseau, France}
\author{Antoine Godard} \affiliation{DSG, ONERA, Université Paris-Saclay, 91120 Palaiseau, France}
\author{Nassim Zahzam} \email{nassim.zahzam@onera.fr}
\affiliation{DPHY, ONERA, Université Paris-Saclay, 91120 Palaiseau, France}%



\date{\today}

\begin{abstract}
We present the simultaneous trapping and cooling of ${}^{85}$Rb, ${}^{87}$Rb and ${}^{133}$Cs in a triple-species magneto-optical trap (MOT). This demonstration is obtained using an all-fibre compact and robust laser system based on telecom 1.5 µm and 2 µm technologies. We characterise the two-body interspecies losses in the double and triple-species MOT. We calculate the two-body interspecies trap-loss coefficient for the ${}^{85}$Rb/${}^{133}$Cs and the ${}^{87}$Rb/${}^{133}$Cs pairs, representing a variation of less than 7 \% in atom number. No losses are observed between ${}^{85}$Rb and ${}^{87}$Rb in our experimental conditions. We find that interspecies interactions inside the triple-species MOT do not prevent trapping and cooling 10$^{8}$ atoms for each species, making our system suitable for future multi-species cold-atom interferometry applications.
\end{abstract}

\maketitle


\section{\label{sec:intro}Introduction}

Cold mixtures of different atomic species have been a topic of growing interest for the last twenty years and have opened up new perspectives in various fields, with many potential applications. For instance, the investigation of ultracold heteronuclear collisions in such mixtures has paved the way to the study of ultracold diatomic molecules \cite{review_diatomic}. Such molecules can be engineered with different atomic species in order to measure fundamental constants \cite{the_nl-eedm_collaboration_measuring_2018,TarbuttYbF2018} or simulate Hamiltonians of many-body physics systems \cite{cornish_quantum_2024}. Equivalence principle tests have also been carried out exploiting the mass difference between two atomic species and the precision of atom interferometry \cite{Hansch_eq_principle, bonnin_simultaneous_2013,tarallo_test_2014,schlippert_quantum_2014,zhou_test_2015,barrett_testing_2022}. Multi-species cold mixtures can be the stage of sympathetic cooling \cite{RbCssympcool,Rb8587sympcool}
which motivated, for instance, the first realisation of a triple-species MOT in 2006 by Tablieger \textit{et al.} using ${}^{87}$Rb, ${}^{6}$Li and ${}^{40}$K \cite{tripleMOT}. Since then, other realisations of three species mixtures cooling have been demonstrated \cite{PhysRevA.84.011601, PhysRevA.105.012816}. However, to our knowledge, no triple-species apparatus is compatible with high-precision inertial measurements based on cold atom interferometry.

The manipulation of different atomic species has yet shown benefits in quantum technologies and metrology \cite{Elliott_2023}. By combining different atomic species, we can expect promising improvements on atomic \cite{dualClock2010} and optical clocks \cite{witkowski_dual_2017}, quantum computing \cite{anand_dual-species_2024}, and cold-atom inertial sensors \cite{tinsley_toward_2024,bonnin_simultaneous_2013}. In particular, a triple-species interferometer offers rich and promising perspectives for improving onboard inertial measurements \cite{bonnin_new_2018}. In this context, the presence of three atomic species also offers new possibilities for interleaved operation, reducing the dead-time \cite{interleaved} or even potentially allows simultaneous 3D acceleration measurement \cite{3Dacc}. This premise has motivated the study of the triple-species MOT using ${}^{85}$Rb, ${}^{87}$Rb and ${}^{133}$Cs, as state-of-the-art cold-atom inertial sensors are based on rubidium and/or caesium \cite{sota_gyro, sota_gravi, Freier_2016}. This triple-species MOT is a first step towards innovative inertial measurements.

Because of the numerous atomic transitions that have to be addressed, the laser system for multi-species operation can quickly become complicated and bulky for onboard applications. In our case, the atomic species ${}^{85}$Rb, ${}^{87}$Rb, and ${}^{133}$Cs allow a compact design for the laser system based on telecom 1.5 µm and 2 µm sources, followed by non-linear up-conversion stages \cite{diboune_multi-line_2017}. Using only 4 laser diodes with phase modulation, we are able to generate notably all the cooling and repumper lines for the three species. Indeed, laser signal at 780 nm can be obtained from one laser diode at 1560 nm for ${}^{85}$Rb/${}^{87}$Rb and 852 nm can be obtained from two laser diodes at 1560 nm and 1878 nm for ${}^{133}$Cs. This laser system is fully fibred and has been designed to be compatible with triple-species interferometry. Our laser system is particularly suitable for on-board applications and could be combined with recent advances made on miniaturisation of cold rubidium and caesium experiments, using grating MOTs, for example \cite{xu_dual_2025}. 

In this paper we report on a simultaneous ${}^{85}$Rb/${}^{87}$Rb/${}^{133}$Cs MOT with a robust and compact laser system. We first describe the experimental apparatus in Section \ref{sec:exp} with the vacuum system, the all-fibre laser system and the imaging system. In Section \ref{sec:methods} we present the loading time measurements in double-species MOTs as well as the triple-species MOT. We evaluate the interspecies trap-loss coefficients in experimental conditions suitable for future cold-atom interferometry and discuss the impact on future multi-species atom interferometry experiments.

\section{\label{sec:exp}Experimental Apparatus}

\subsection{\label{sec:level2}Vacuum system}

\begin{figure}[h]
\includegraphics[width=7cm]{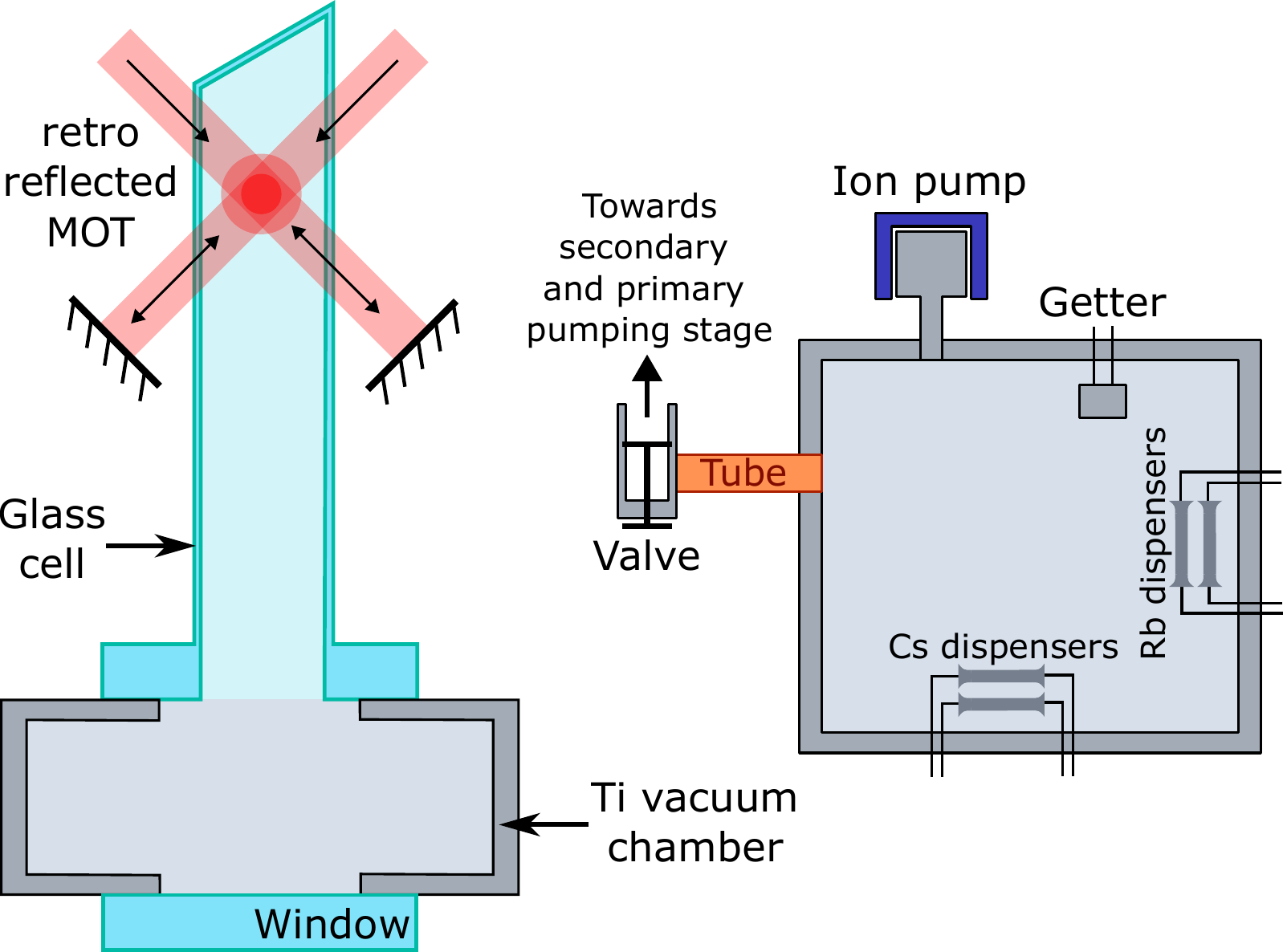}
\caption{\label{fig:vacuum} Schematic of the UHV cell. Sectional side-view (left) and top-view of the vacuum chamber (right). The region of the retro-reflected MOT is indicated on the side-view. Two pairs of dispensers generate  Rb and Cs vapour, respectively. A getter and an ion pump maintain a UHV of $\sim10^{-9}$ mbar. A copper pinch-off tube links the chamber to a valve that can be connected to a primary and secondary pumping bench if needed.}
\end{figure}

The MOTs are produced in the top-part of a glass cell, mounted on a Ti vacuum chamber (Figure \ref{fig:vacuum}). An ion pump and a getter maintain an ultra-high vacuum (UHV) of $\sim10^{-9}$ mbar. We use commercial dispensers to deliver the two natural isotopes of rubidium ${}^{85}$Rb/${}^{87}$Rb and ${}^{133}$Cs. Each pair of dispensers is typically powered with 2.5 A, generating a vapour of about $10^{8}$ atoms/$\mathrm{cm}^3$ for each of the three species. 

\subsection{\label{sec:level2}Laser system}

\begin{figure}[h]
\includegraphics[width=8.5cm]{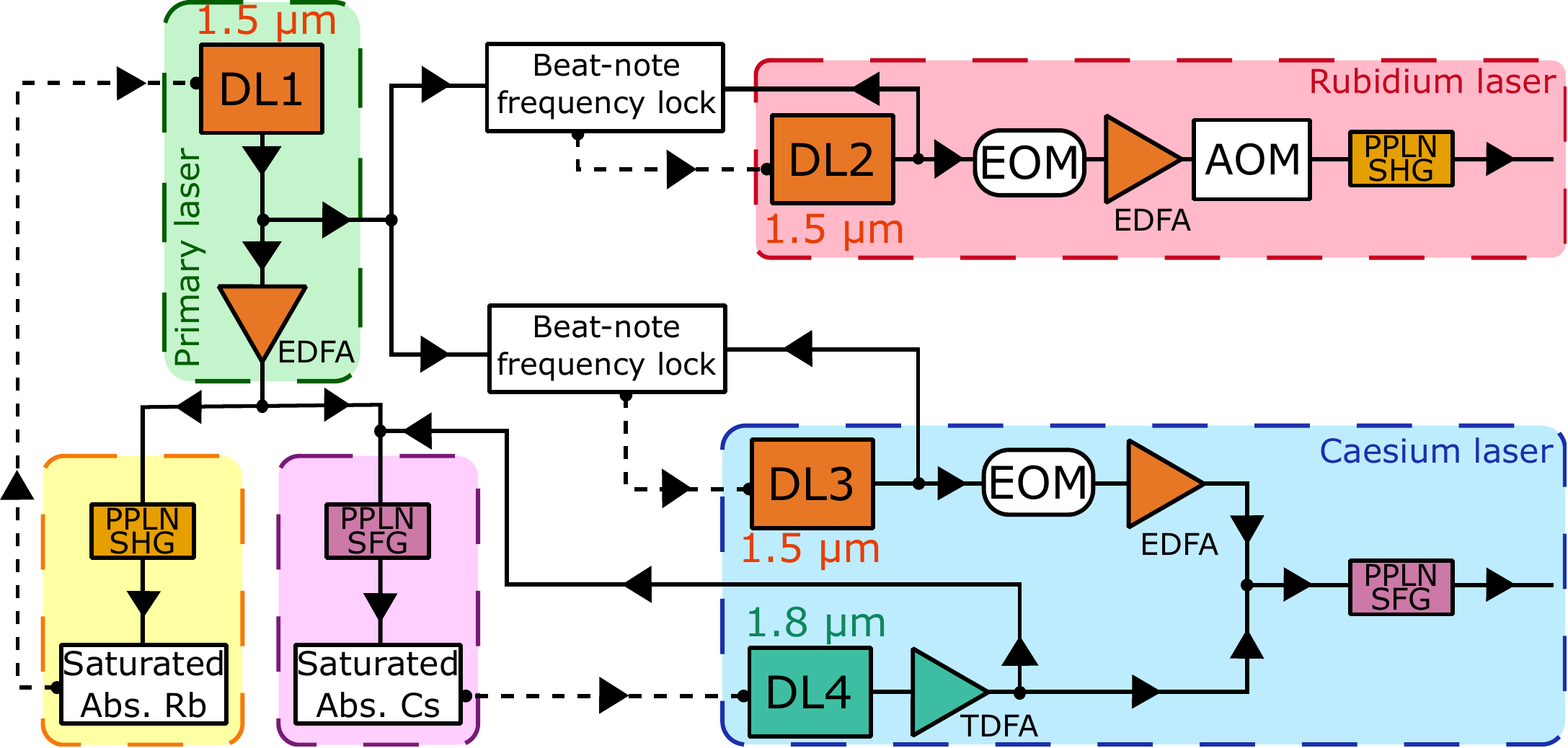}
\caption{\label{fig:laser} Simplified scheme of the laser system for the triple-species MOT. Each 1560 nm diode (colour-coded in orange) is amplified by an erbium-doped fibre amplifier (EDFA). The diode at 1878 nm is amplified by a thulium-doped fibre amplifier (TDFA), in green. The PPLN SHG and SFG are respectively used for second-harmonic generation and sum frequency generation.  Phase modulation is generated with the EOMs operating before the up-conversion stages. The dashed lines represent the feedback signal of the laser frequency lock-in. The acousto-optic modulator (AOM) on the rubidium track serves as a junction between the MOT track and a laser track dedicated to 780 nm/852 nm interferometry. The laser track for interferometry is not represented in this figure.}
\end{figure}

The laser system is based on telecom 1.5 µm and 2 µm technologies, followed by frequency up-conversion stages. Second harmonic generation (SHG) at telecom wavelength 1560 nm enables to obtain 780 nm for Rb \cite{carraz_compact_2009}, while sum frequency generation (SFG) between 1560 nm and 1878 nm sources enables to obtain 852 nm for Cs \cite{diboune_multi-line_2017, exail_laser}. The setup is fully fibred and relies on only four tunable laser diodes for the entire triple-species MOT operation. A simplified scheme of the laser system is presented in Figure \ref{fig:laser}. A first 1560 nm laser diode DL1 is frequency-doubled in a periodically-poled lithium niobate crystal (PPLN) and frequency locked on the cross-over transition $F = 3 \leftrightarrow 3 \times 4$ of ${}^{85}\mathrm{Rb}$ using saturated absorption spectroscopy. This laser serves as a frequency reference. A beat-note lock system allows to lock the other laser diodes at 1560 nm to the primary laser DL1. The beat-note is controlled with voltage-controlled oscillators, allowing fine control of the detuning of the lasers.\\
To generate the appropriate optical signal for a rubidium MOT, a 1560 nm laser diode DL2 is locked on DL1 with a beat-note of 650.8 MHz, such that, once frequency doubled, the frequency of DL2 corresponds to the cooling line of ${}^{87}\mathrm{Rb}$. The cooling line corresponds to the transition $F = 2 \leftrightarrow F' = 3$ being red-detuned by 2.5 $\Gamma_{\mathrm{Rb}}$, with $\Gamma_{\mathrm{Rb}} \approx 2\pi \times 6$ MHz, the natural linewidth of this Rb atomic transition. To generate the cooling line of ${}^{85}\mathrm{Rb}$ and repumping lines for both ${}^{87}\mathrm{Rb}$ and ${}^{85}\mathrm{Rb}$, we use an electro-optic modulator (EOM) operating at 1560 nm. The cooling line of ${}^{85}\mathrm{Rb}$, red-detuned by 2.5 $\Gamma_{\mathrm{Rb}}$ from $F = 3 \leftrightarrow F' = 4$, and the repumping lines of ${}^{85}\mathrm{Rb}$ $F = 2 \leftrightarrow F' = 3$ and ${}^{87}\mathrm{Rb}$ $F = 1 \leftrightarrow F' = 2$ are generated through phase modulation at respectively 1.12 GHz, 6.58 GHz and 2.93 GHz. The microwave signals for phase modulation can be controlled in frequency (over $\approx 20 \Gamma_{\mathrm{Rb}}$) to tune the laser frequency, as well as in power to adjust the relative line intensities. 

\begin{figure*}
\includegraphics[width=8.7cm]{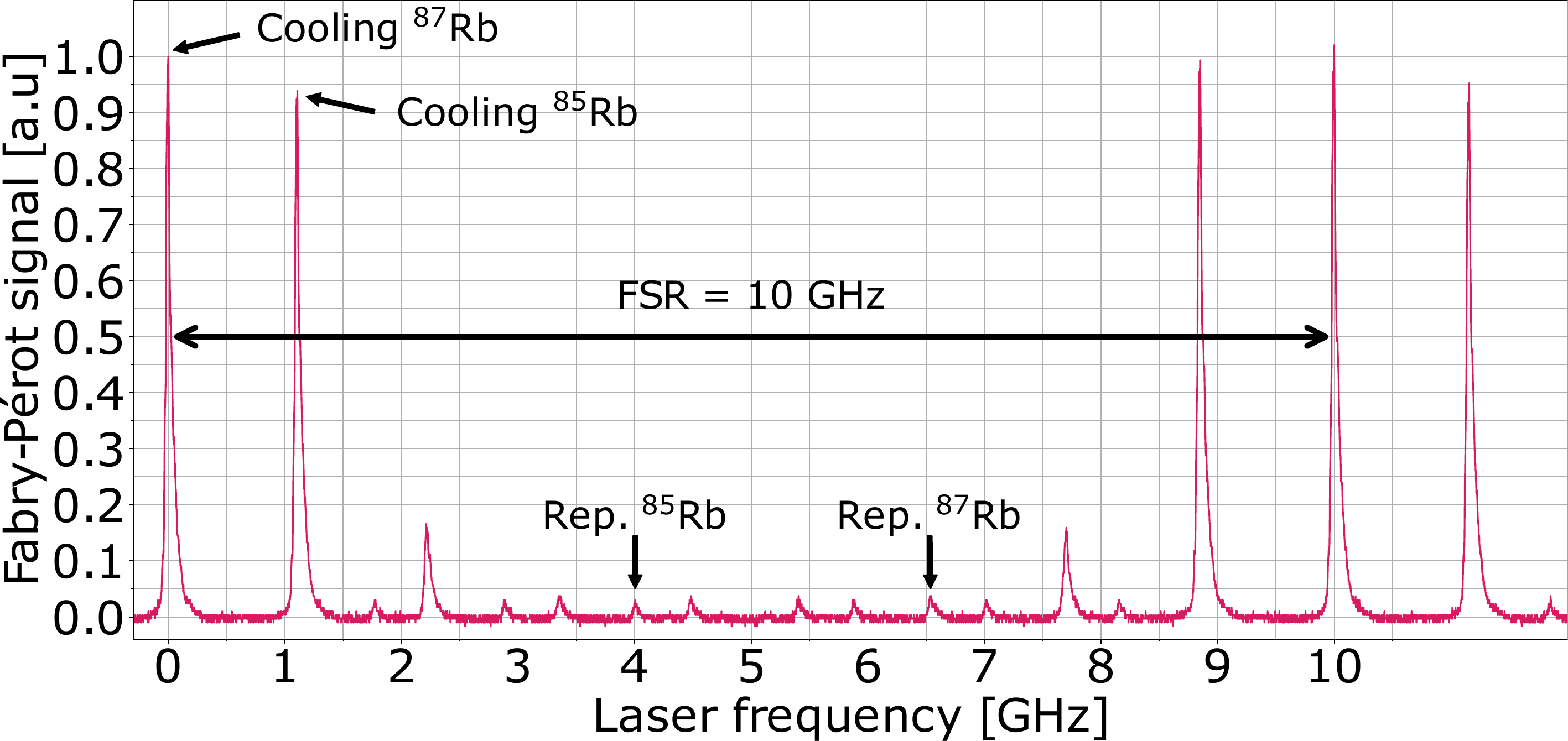}%
 \quad
\includegraphics[width=8.7cm]{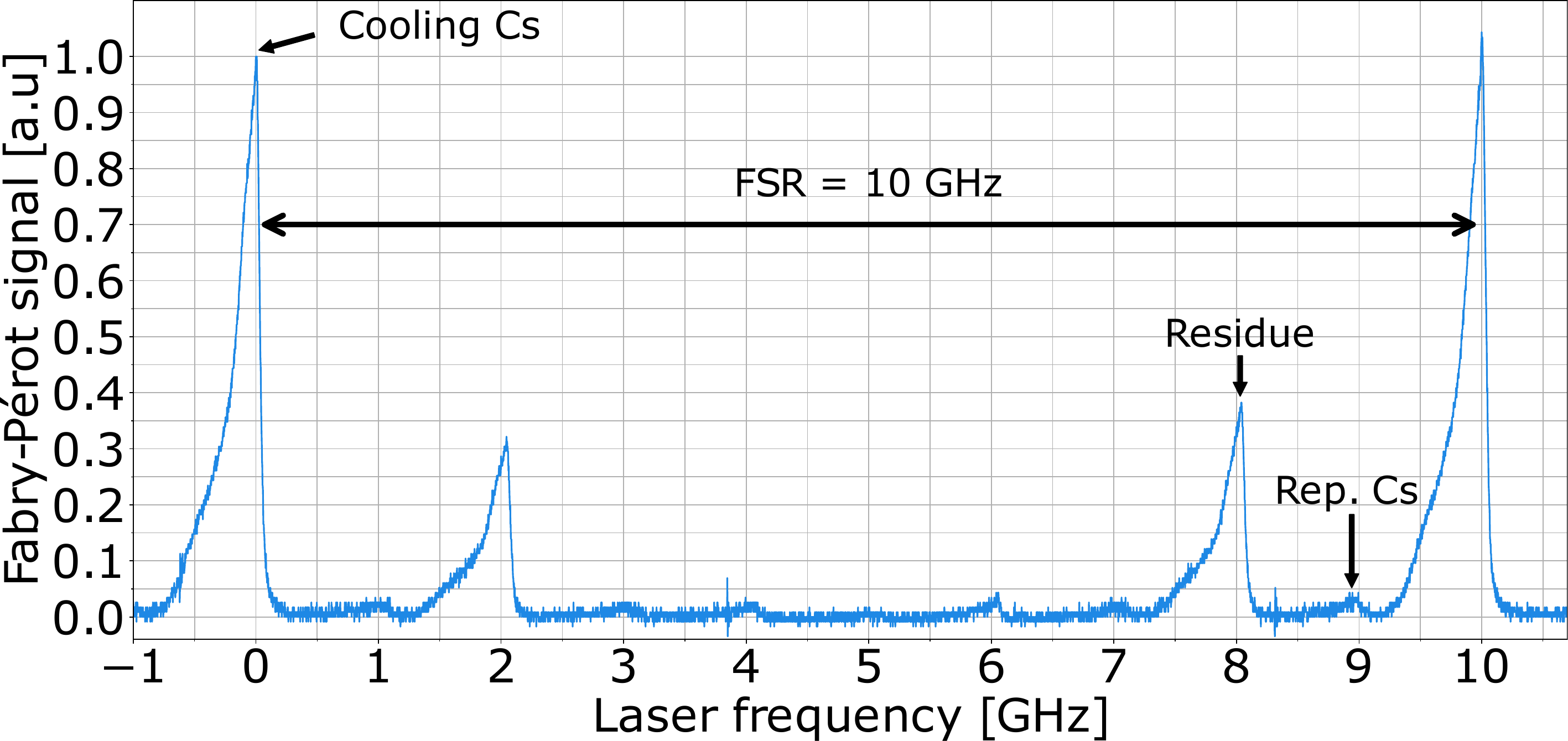}%
 \quad
 \caption{\label{FP}
  Fabry-Pérot spectra of the MOT laser beams at 780 nm (left) and at 852 nm (right). The free spectral range (FSR) is 10 GHz. Each of the relevant lines for cooling and repumping is identified as well as the residue for the 852 nm spectrum. The other peaks are other orders created by the phase modulation.} Misalignment of the 852 nm Farby-Pérot widens the peaks.
\end{figure*}

For caesium, a 1560 nm laser diode DL3 and a 1878 nm laser diode DL4 are frequency-summed to obtain 852 nm \cite{diboune_multi-line_2017}. In order to lock DL4, a sample of the DL4 laser is frequency-summed with the reference laser DL1. DL4 is then locked on the cross-over transition $F = 3 \leftrightarrow 3 \times 4$ of ${}^{133}\mathrm{Cs}$ using saturated absorption spectroscopy. Finally, DL3 is locked on the beat-note frequency of 8915 MHz between DL3 and DL1. The frequency of the resulting DL3 + DL4 signal equals the cooling line of ${}^{133}\mathrm{Cs}$. The cooling line of caesium is the transition $F = 4 \leftrightarrow F' = 5$  red-detuned by 2.5 $\Gamma_{\mathrm{Cs}}$, with $\Gamma_{\mathrm{Cs}} \approx  2\pi \times 5$ MHz, the linewidth of the atomic transition for Cs. The repumping line for caesium $F = 3 \leftrightarrow F' = 4$ is generated 8.9 GHz away from the cooling line via phase modulation of DL3 at 1560 nm.

The MOT beams at 780 nm and 852 nm are then each split into 3 nearly balanced beams through a combination of 50/50 and 70/30 fibred components, and recombined with a 780/852 nm wavelength division multiplexer. The resulting 3 beams, containing the six laser lines described earlier, are then sent in a retro-reflected MOT optical set-up. Each beam is collimated with a lens of focal length 75 mm, resulting in a  25 mm diameter beam. Using a set of quarter-wave plates, the MOT is produced in the $\sigma^+ -\sigma^-$ polarisation configuration. A pair of magnetic coils generates a quadrupole trapping field with a gradient of 15 G/cm.

We decided to operate the multi-species MOT in comparable conditions for each species to match with the experimental conditions for future cold-atom interferometry. Therefore, the microwave signals were adjusted such that the intensity $I$ in the cooling line corresponds to $1.3 \times I_{sat}$, with $I_{sat}$ the saturation intensity of the transition. We define the saturation parameter at resonance as $s = \mathrm{I}/\mathrm{I}_{sat}$. This condition translates to a total optical power of 140 mW at 780 nm, equally shared in the cooling lines of 87 and 85, and 52 mW at 852 nm in the cooling line of caesium. All repumping lines are resonant, and their power corresponds to about 3\% of the corresponding cooling lines, for each of the 3 species. The spectra of the MOT laser at 780 nm and at 852 nm are measured with scanning Fabry-Pérot interferometers, as shown in Figure \ref{FP}. Note that the spectrum at 852 nm features an additional line 8 GHz from the cooling line. This frequency is not resonant with any of the transitions and originates from a spurious frequency in the microwave system. A filter is still to be added to mitigate this 8 GHz frequency component.

\subsection{\label{sec:level2}MOT imaging system}

\begin{figure}[h]
\includegraphics[width=7cm]{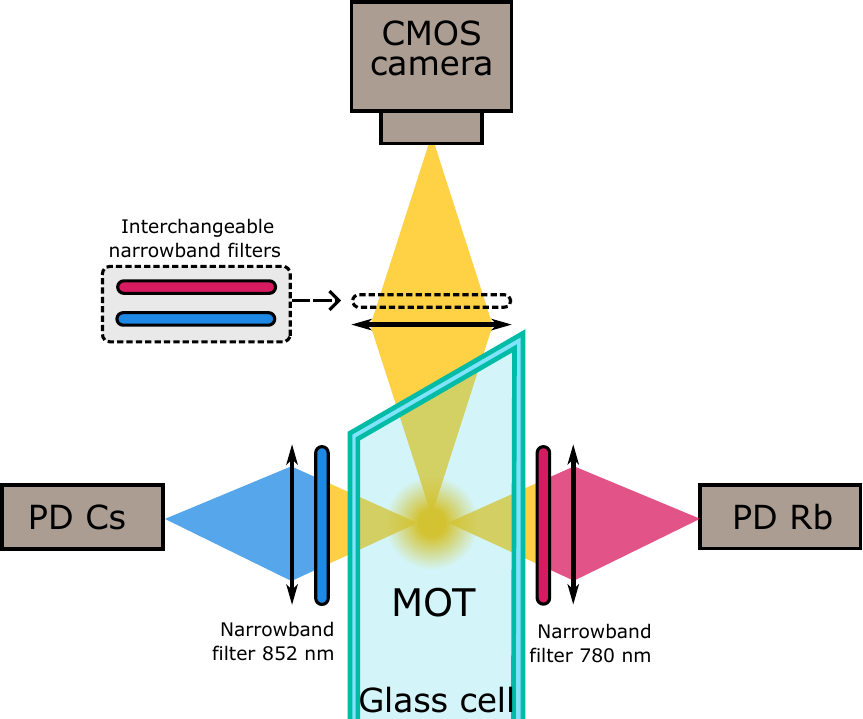}
\caption{\label{fig:imaging} Schematic of the MOT imaging system. The CMOS camera allows spatial imaging of the MOT and can be triggered with the sequence. A set of photodiodes (PD) monitors the fluorescence of the MOT. Narrow band filters can be placed in the imaging system to independently collect the fluorescence at 780 nm, which is the sum of the fluorescence from ${}^{85}\mathrm{Rb}$ and ${}^{87}\mathrm{Rb}$, and the fluorescence at 852 nm.}
\end{figure}

The imaging system for the MOT is shown in Figure \ref{fig:imaging}. The fluorescence of the MOT is captured using a pair of photodiodes on opposite sides of the vacuum cell, with each having its own narrow band $\pm$ 5 nm filter at 780 nm and 852 nm, respectively. It allows for independent monitoring of Rb and Cs. The fluorescence spectrum of ${}^{85}\mathrm{Rb}$ and ${}^{87}\mathrm{Rb}$ being a few GHz apart, we are not able to discriminate between the fluorescence of both isotopes with our current setup. A CMOS camera is installed on top of the cell and images the MOT from above. A set of the same narrow-band filters can be installed in a steady position on top of the mount in order to image at both wavelengths independently.

To estimate the density of the cloud, we fit a 2D Gaussian function on the data from the CMOS camera, with the $1/e^2$ radii $w_x$, $w_y$, and the position of the centre of the cloud $(x;y)$ as free parameters. We observe elliptic shapes of the clouds in the radial plane of the coil, especially for caesium. The latter can be explained by an imbalance in the power of the 3 MOT beams at 852 nm, already observed during the characterisation of the fibred recombination. The MOTs have a typical radius of $w_x \approx w_y \approx 1$ mm and contain around $10^{8}$ atoms for each species, giving an average density of about $3\times10^{10}$ $\mathrm{atoms/cm}^3$. Note that we have to extrapolate the third dimension of the cloud because the imaging system only provides a 2D map of the MOT. We decided to take the mean value of $w_x$ and $w_y$ to approximate the cloud radius along the vertical axis $w_z$.

\section{\label{sec:methods}Methods and results}

To observe the loading of a MOT, we operate a sequence in which we switch on and off the repumping line with the EOM. In order to study the MOTs of the different species in similar conditions, the optical power of each cooling line at 780 nm is adjusted and kept constant while switching on and off the repumper modulation frequency.

For each species, we can generate single-species MOTs of around $1.5\times10^{8}$ atoms of ${}^{133}\mathrm{Cs}$, $1.2\times10^{8}$ atoms of ${}^{87}\mathrm{Rb}$, and $1.7\times10^{8}$ atoms of ${}^{85}\mathrm{Rb}$ atoms, with a loading time of around 159 ms, 162 ms and 168 ms respectively for ${}^{133}\mathrm{Cs}$, ${}^{85}\mathrm{Rb}$ and ${}^{87}\mathrm{Rb}$. The ratio ${}^{85}\mathrm{Rb}$/${}^{87}\mathrm{Rb}$ is 60/40.

We started by investigating the two-species MOTs for each combination of species that we could generate. Each measurement is split in 2 sequences, in which we load atom A in a MOT B that is already established, and conversely for the second sequence. We made sure that the laser at 780 nm did not influence the trapping of caesium, and conversely with the laser at 852 nm for rubidium. Thus, we can rule out any significant impact from lightshifts on the MOT. In order to validate the MOT set-up, we evaluate the interspecies trap-loss coefficients in the multi-species MOT.

\subsection{\label{sec:Double-species MOT}Double-species MOT}
\begin{figure*}
\includegraphics[width=10.7cm]{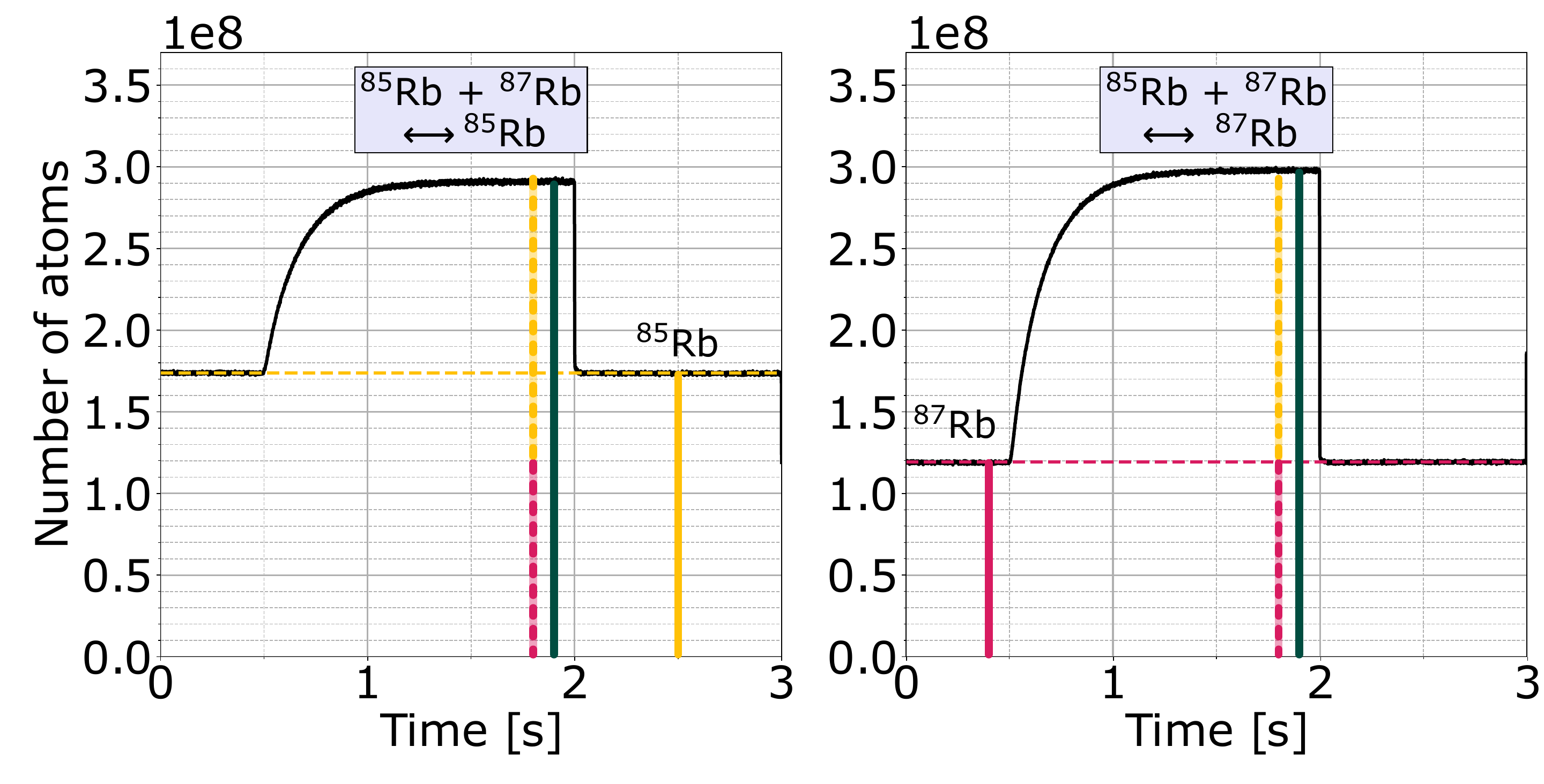}
\caption{\label{fig:8587} Double-species MOT loading curves from ${}^{85}\mathrm{Rb}$ + ${}^{87}\mathrm{Rb}$ MOTs. The first plot corresponds to a sequence switching on and off the repumper of ${}^{87}\mathrm{Rb}$. The second plot corresponds to a sequence switching on and off the repumper of ${}^{85}\mathrm{Rb}$. The number of atoms in each of the single MOTs is represented as solid bars, red for ${}^{87}\mathrm{Rb}$ and yellow for ${}^{85}\mathrm{Rb}$. The measurements are reported as dashed bars, stacked on top of each other, to compare with the number of atoms in the double rubidium MOT, which atom number is represented as dark green bars.}
\end{figure*}

Comparing the number of atoms in the single-species MOT and in the multi-species MOT can tell us about the interactions between the species. If interspecies losses occur, we expect to have fewer atoms in the multi-species MOT compared to the sum of the number of atoms in each single-species MOT. Figure \ref{fig:8587} shows two 3-second sequences in which we alternately load a double ${}^{85}\mathrm{Rb} + {}^{87}\mathrm{Rb}$ MOT and a single ${}^{85}\mathrm{Rb}$ or ${}^{87}\mathrm{Rb}$ MOT. The bars represent the number of atoms in the different MOTs: the single ${}^{85}\mathrm{Rb}$ MOT (yellow bar), the single ${}^{87}\mathrm{Rb}$ MOT (red bar), and the double ${}^{85}\mathrm{Rb} + {}^{87}\mathrm{Rb}$ MOT (dark green bar). Comparing the number of atoms in the double MOT with the sum of the number of atoms in ${}^{87}\mathrm{Rb}$ and ${}^{85}\mathrm{Rb}$ single-species MOT (red and yellow dashed bars put end to end), we notice that both quantities are equal, at a level of uncertainty estimated to 3\%. This uncertainty comes from the background vapour fluctuation between the measurements. This equality indicates that the presence of a species on the other induces no significant losses. Another confirmation is the shape of the loading curve, as the de-load of the double-species MOT falls flat. If losses were induced, some reload could be observed, as we will see later in the triple-species MOT. Indeed, due to the suppression of collisions between ${}^{85}\mathrm{Rb}$ and ${}^{87}\mathrm{Rb}$, the MOT of the remaining species would experience some reload. It is, however, not the case in Figure \ref{fig:8587}.

\begin{figure*}
\includegraphics[width=10.7cm]{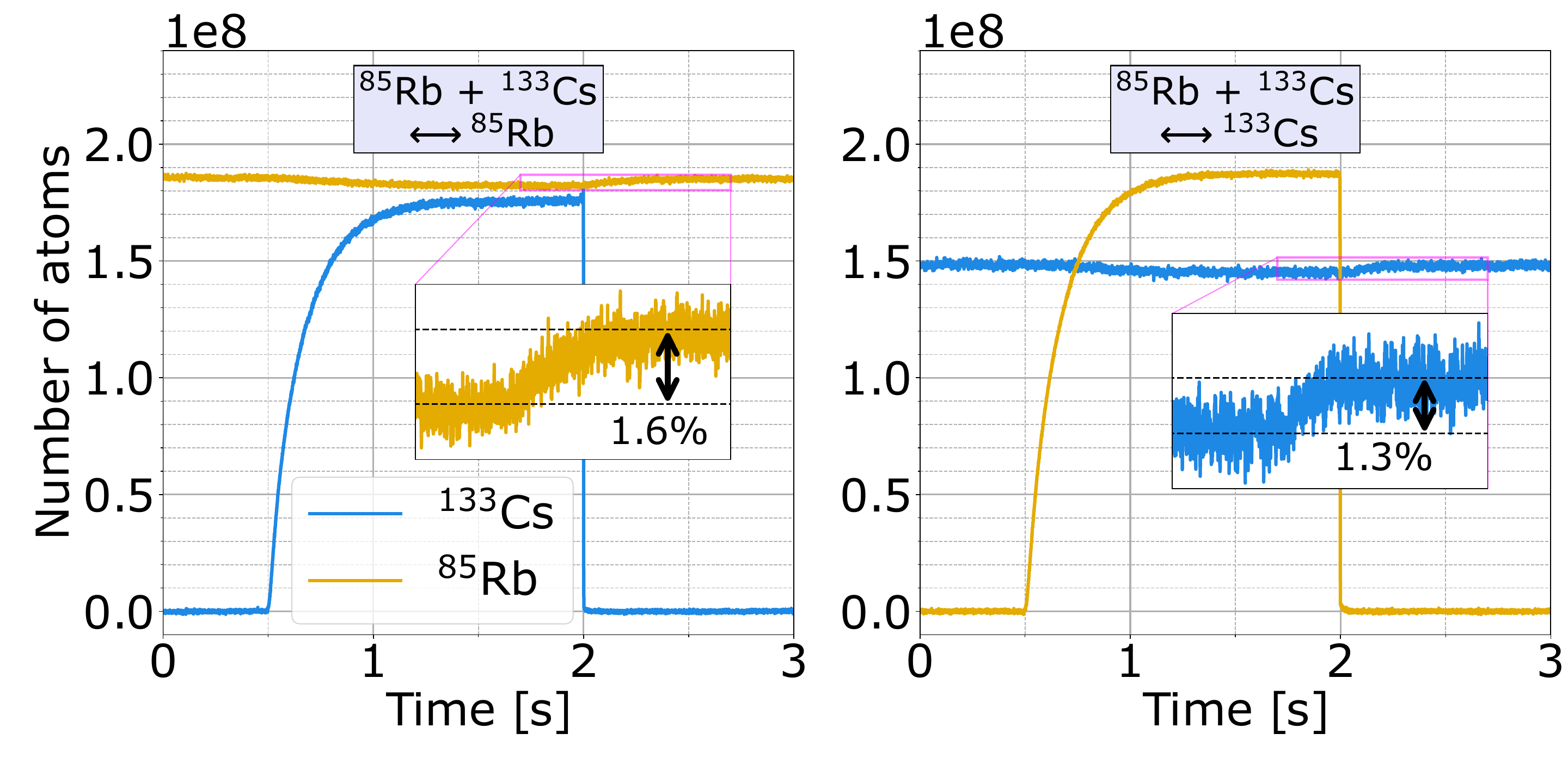}
\caption{\label{fig:85Cs} Double-species MOT loading curves from ${}^{85}\mathrm{Rb}$ + ${}^{133}\mathrm{Cs}$ MOTs. The first plot corresponds to a sequence switching on and off the repumper of ${}^{133}\mathrm{Cs}$. The second plot corresponds to a sequence switching on and off the repumper of ${}^{85}\mathrm{Rb}$. The number of ${}^{133}\mathrm{Cs}$ is in blue, and the number of ${}^{85}\mathrm{Rb}$ is in yellow. The inset shows the variation between the double and the single MOTs of $\sim 1-2\%$, interpreted as arising from 2-body collisions between ${}^{133}\mathrm{Cs}$ and ${}^{85}\mathrm{Rb}$. }
\end{figure*}

Figure \ref{fig:85Cs} shows a similar experiment with ${}^{85}\mathrm{Rb}$ and ${}^{133}\mathrm{Cs}$. Since we have two separate photodiodes to monitor the MOTs (Figure \ref{fig:imaging}), we can independently observe both species in the double MOT. When ${}^{133}\mathrm{Cs}$ is added, the number of ${}^{85}\mathrm{Rb}$ atoms decreases by a 1.6 \%. The same observation holds for the second part of Figure \ref{fig:85Cs}, adding ${}^{85}\mathrm{Rb}$ into an already established ${}^{133}\mathrm{Cs}$ MOT. The number of ${}^{133}\mathrm{Cs}$ atoms decreases by 1.3 \%.

\begin{figure*}
\includegraphics[width=10.7cm]{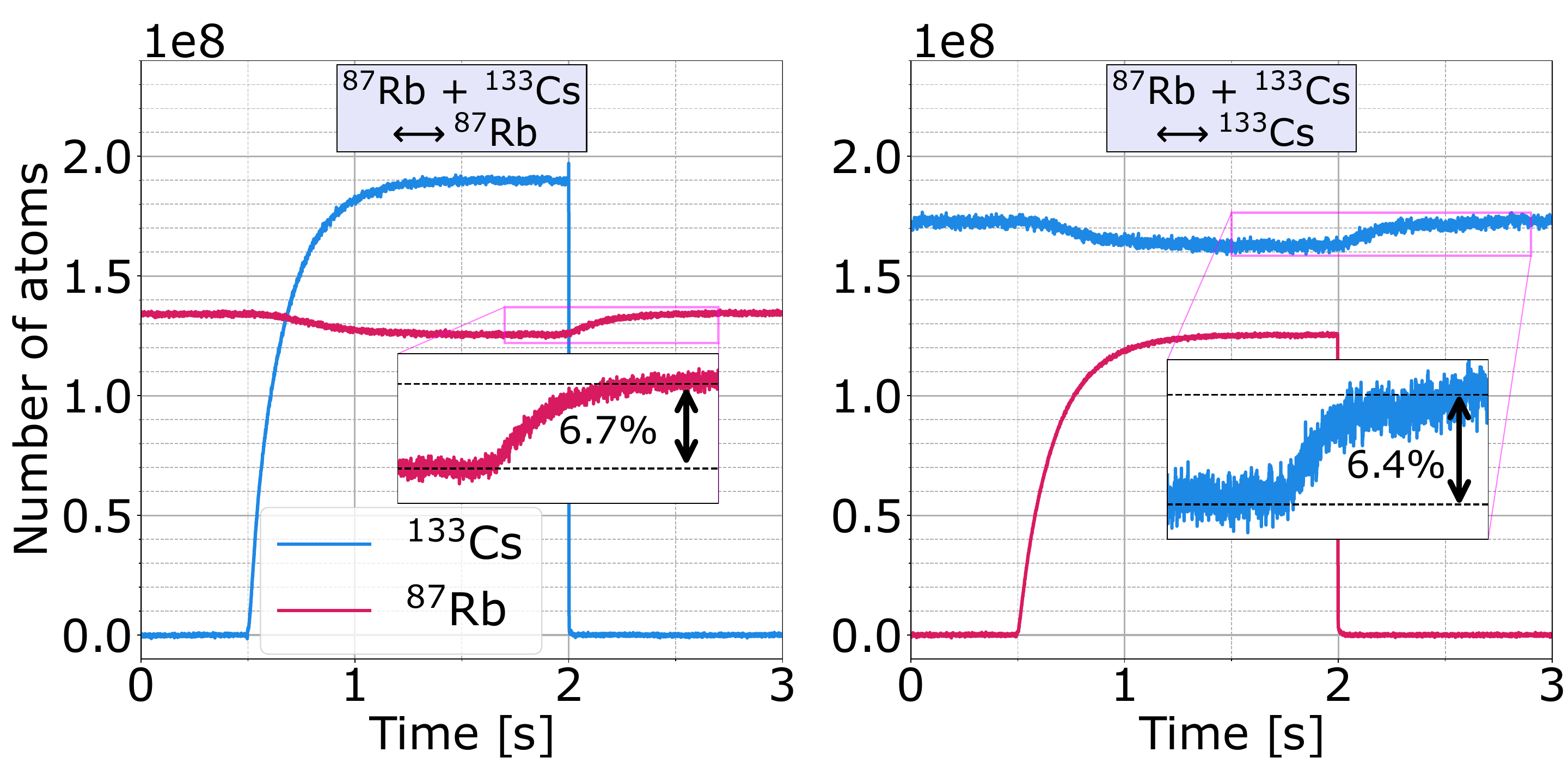}
\caption{\label{fig:87Cs} Double-species MOT loading curves from ${}^{87}\mathrm{Rb}$ + ${}^{133}\mathrm{Cs}$ MOTs. The first plot corresponds to a sequence switching on and off the repumper of ${}^{133}\mathrm{Cs}$. The second plot corresponds to a sequence switching on and off the repumper of ${}^{87}\mathrm{Rb}$. The number of ${}^{133}\mathrm{Cs}$ is in blue, and the number of ${}^{87}\mathrm{Rb}$ is in red. The inset shows the variation between the double and the single MOTs of $\sim 6-7\%$, interpreted as arising from 2-body collisions between ${}^{133}\mathrm{Cs}$ and ${}^{87}\mathrm{Rb}$.}
\end{figure*}

For the ${}^{87}\mathrm{Rb}$ + ${}^{133}\mathrm{Cs}$ study, even more significant variations can be observed (Figure \ref{fig:87Cs}). When a ${}^{133}\mathrm{Cs}$ MOT is loaded into a ${}^{87}\mathrm{Rb}$ MOT, the number of ${}^{87}\mathrm{Rb}$ atoms decreases by 6.7\% change. In a similar case, adding ${}^{87}\mathrm{Rb}$ into the ${}^{133}\mathrm{Cs}$ MOT leads to a decrease of 6.4\%.

Qualitatively, in the current set of parameters, 85-87 trap-loss collisions appear to be negligible in the double-species MOT, whilst they have a measurable impact in the 85-Cs and 87-Cs double MOTs. Interspecies collision-induced losses are stronger between 87-Cs than 85-Cs. Some quantitative analysis will corroborate this observation in Section \ref{sec:traploss}.

\subsection{\label{sec:Triple-species MOT}Triple-species MOT}

\begin{figure}[h]
\includegraphics[width=8.6cm]{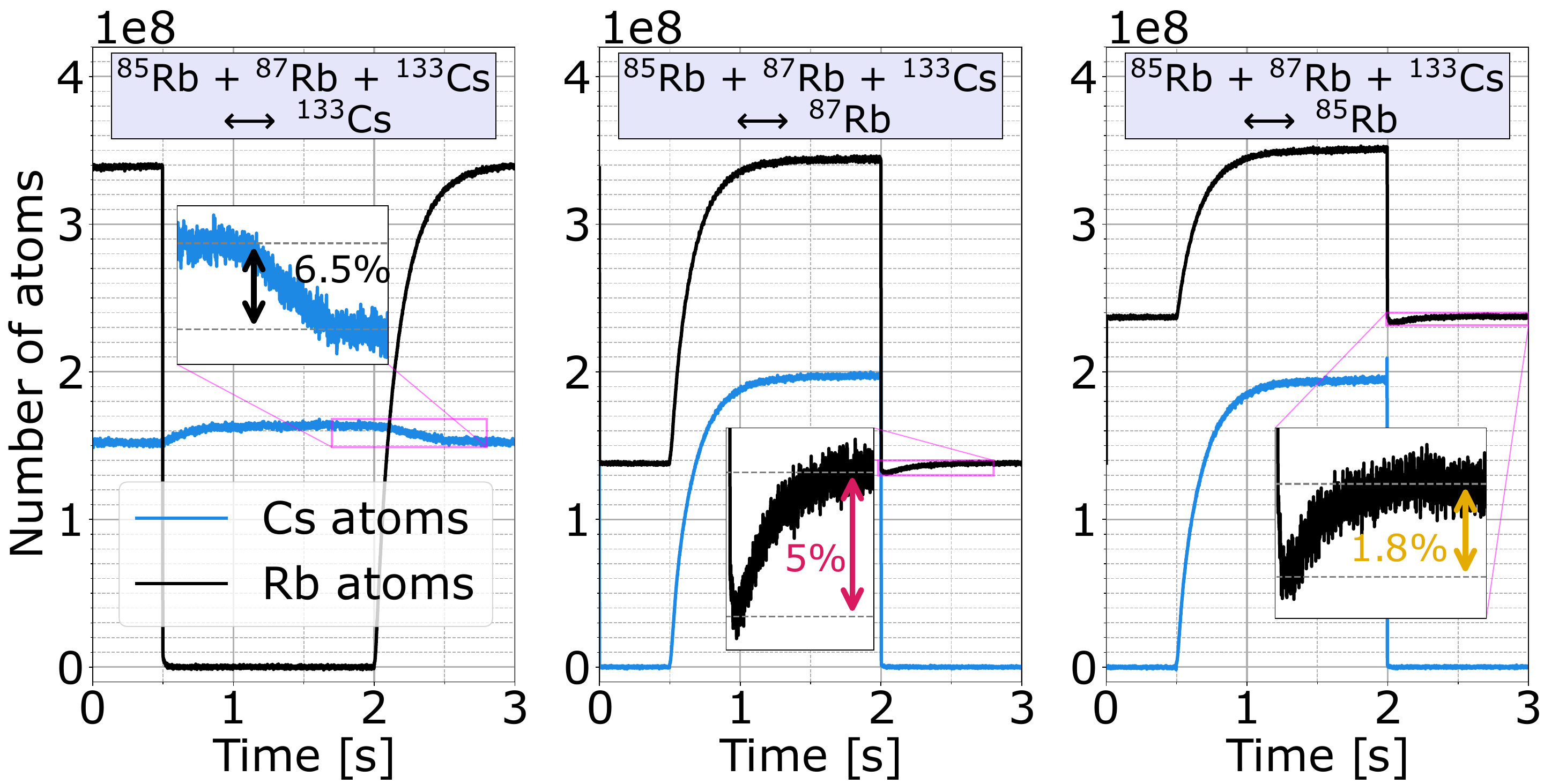}
\caption{\label{fig:triple} Triple-species MOT ${}^{87}\mathrm{Rb}$+${}^{85}\mathrm{Rb}$+${}^{133}\mathrm{Cs}$ loading curves. The three plots correspond to sequences switching on and off the repumpers to alternate between a triple MOT and a MOT of ${}^{133}\mathrm{Cs}$ (left), ${}^{87}\mathrm{Rb}$ (middle) or ${}^{85}\mathrm{Rb}$ (right). The number of ${}^{133}\mathrm{Cs}$ is in blue, and the number of rubidium, ${}^{87}\mathrm{Rb}$ and/or ${}^{85}\mathrm{Rb}$ is in black. We observe a variation the number of atoms of 6.5\% for ${}^{133}\mathrm{Cs}$, 5\% for ${}^{87}\mathrm{Rb}$ and 1.6\% for ${}^{85}\mathrm{Rb}$.}
\end{figure}

In the triple-species MOT, atom numbers of $1.3 \times 10^{8}$ ${}^{87}\mathrm{Rb}$, $2.3 \times 10^{8}$ ${}^{85}\mathrm{Rb}$ and $1.9 \times 10^{8}$ ${}^{133}\mathrm{Cs}$ were simultaneously trapped. We note trap-loss effects by switching from a triple-species MOT to a single-species MOT (Figure \ref{fig:triple}). After switching off the triple-species MOT into a single-species MOT, we observe some reloading of the MOT, where the number of atoms slightly increases. This reload corresponds to the difference in the number of atoms in the multi-species MOT and in the single species MOT that is caused by losses occurring in the multi-species MOT only. This loss is attributed to 2-body interspecies collisions that still mainly occur between Rb and Cs. Since no collisional effects were observed between ${}^{85}\mathrm{Rb}$ and ${}^{87}\mathrm{Rb}$ in the double-species MOT, we assume that they will also not be observed in the triple-species MOT. Three-body collisions between the three species could potentially be observed and measured with our setup, however they are likely to be negligible given the relatively low atomic density of the MOT. To verify this assumption, an upper bound of the three-species coefficient will be calculated in the next section. Moreover, we observe in Figure \ref{fig:triple} a change in the isotopic proportions of rubidium within the triple-species MOT. Indeed, the ratio of ${}^{85}\mathrm{Rb}$/${}^{87}\mathrm{Rb}$ is 65/35 in the triple MOT instead of 60/40 that was observed in Section \ref{sec:Double-species MOT}, in the absence of ${}^{133}\mathrm{Cs}$. This ratio difference stems from the trap-loss interaction between ${}^{87}\mathrm{Rb}$ and ${}^{133}\mathrm{Cs}$ being higher compared to ${}^{85}\mathrm{Rb}$ and ${}^{133}\mathrm{Cs}$. This results in more losses for ${}^{87}\mathrm{Rb}$, and therefore a lower proportion of ${}^{87}\mathrm{Rb}$ in the triple-species MOT.

\subsection{\label{sec:traploss}Interspecies trap-loss collisions}

The loading of a species $i$ inside a multi-species MOT containing three atomic species that we name $i$, $j$, and $k$ can be modelled by the following rate equation:

\begin{eqnarray} \label{eq:1}
\frac{dN_i^{(3)}}{dt}= && L_i-\gamma_iN_i^{(3)}-\beta_i\iiint\limits_{V_i}{{n_i^{(3)}(\vec{r})}^2dV_i}\nonumber\\
&& -\beta_{i,j}\iiint\limits_{V_i}{n_i^{(3)}(\vec{r})n_j^{(3)}(\vec{r})dV_i}\\
&&-\beta_{i,k}\iiint\limits_{V_i}{n_i^{(3)}(\vec{r})n_k^{(3)}(\vec{r})dV_i}\nonumber
\end{eqnarray}
Where $N_{i}^{(3)}$ is the number of atoms of $i$ in the triple-species MOT, $L_{i}$ is the loading rate for species $i$, and $\gamma_{i}$ is the loss rate due to collisions between the hot background vapour and cold atoms of $i$. $\beta_{i}$ is the loss rate coefficient for binary collisions between atoms of the same species $i$ within the cloud of volume $V_{i}$ and of density $n_{i}^{(3)}(\vec{r})$. We will denote the average density of the cloud as $n_{i}^{(3)}$. The superscript (3), (2) or (1) indicates whether the quantity is evaluated in a triple, double, or single species MOT. Due to the presence of the species $j$ and $k$, we also need to consider collisions between the cold atoms of different species. $\beta_{i,j}$ and $\beta_{i,k}$ denote the loss rate coefficient for binary collisions between different species, involving, respectively, cold atoms and  $i$ and $j$, and cold atoms of $i$ and $k$. The order of indices in $\beta_{i,j}$ is relevant, since it describes the losses induced by the collisions, and not the collision rate itself. The convention we use for $\beta_{i,j}$ is the following: the first index $i$ stands for the species being lost due to the presence of the species $j$ indicated by the second index \cite{dutta_interspecies_2014}. By construction, here $i \neq j$. 

In practice, we will measure the interspecies trap-loss coefficients using the data from the double-species MOTs, which reduces Eq. \ref{eq:1} to only 2 species $i$ and $j$. We assume that the density of the MOT remains constant during loading \cite{mancini_trap_2004}, which is verified for MOT containing more than $10^5$ atoms \cite{Cst_density}. The loading Equation \ref{eq:1} can then be simplified in the following expression for a double-species MOT:

\begin{eqnarray} \label{eq:2}
\frac{dN_i^{(2)}}{dt}=L_i-\gamma_iN_i^{(2)}-\beta_in_i^{(2)}N_i^{(2)}-\beta_{i,j}n_j^{(2)}N_i^{(2)}
\end{eqnarray}
Which can now be solved analytically as:
\begin{eqnarray} \label{eq:3}
N_i^{(2)}\left(t\right)=&{N_i^{(2)}}^\infty\ (\ 1\ -e^{-t/\tau_i^{(2)}}\ )\nonumber \\
{N_i^{(2)}}^\infty=&\frac{L_i}{\gamma_i+\beta_in_i^{(2)}+\beta_{i,j}n_j^{(2)}} \\
\tau_i^{(2)}=&\frac{1}{\gamma_i+\beta_in_i^{(2)}+\beta_{i,j}n_j^{(2)}} \nonumber
\end{eqnarray}
Where ${N_i^{(2)}}^\infty$ is the stationary number of atoms $i$ in the double-species MOT, and $\tau_i$ is the characteristic loading time. In a single species MOT, the number of atoms in the steady state is given by Equation \ref{eq:3} with $\beta_{i,j} = 0$. Consequently, one can arrange the expressions of the number of atoms in the single MOT ${N_i^{(1)}}^\infty$ and in the double MOT ${N_i^{(2)}}^\infty$, and obtain the following expression of $\beta_{i,j}$:

\begin{eqnarray} \label{eq:4}
\beta_{i,j}=\frac{1}{{n_j^{(2)}}\tau_i^{(1)}}\left(\frac{{N_i^{(1)}}^\infty}{{N_i^{(2)}}^\infty}-1\right)
\end{eqnarray}

\begin{figure*}[ht]
\includegraphics[width=17.5cm]{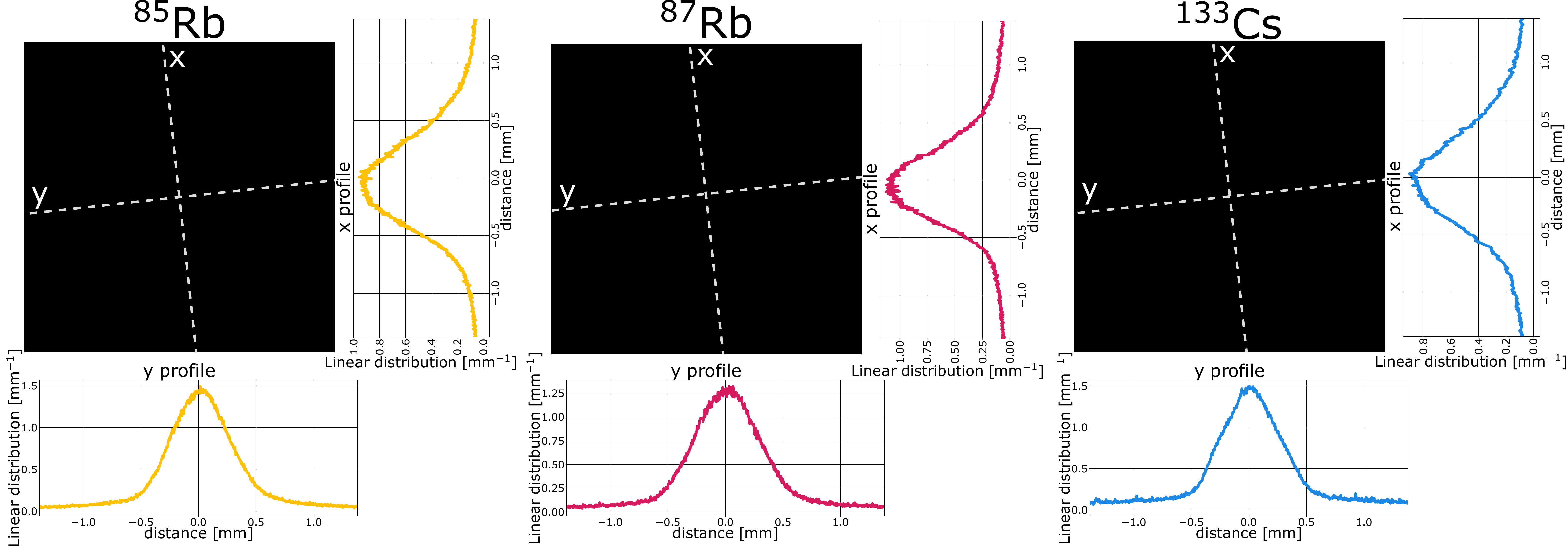}
\caption{\label{fig:density} From left to right: false colour fluorescence of the ${}^{85}\mathrm{Rb}$ (yellow), ${}^{87}\mathrm{Rb}$ (red) and ${}^{133}\mathrm{Cs}$ (blue) simultaneously trapped in the triple species MOT. Each image has its background subtracted, and the three of them are layered on one another, alternating the species on top. The signal from ${}^{87}\mathrm{Rb}$ and ${}^{85}\mathrm{Rb}$ at 780 nm is identified by triggering the CMOS camera, imaging the MOT when the fluorescence drops right after the switch from a 3-species MOT to a single ${}^{87}\mathrm{Rb}$ (respectively ${}^{85}\mathrm{Rb}$) MOT. The major and minor axes of the 2D Gaussian fit are represented as dashed lines in white. The axes of the Gaussian of the three species coincide within 2$^{\circ}$. The profiles along $x$ and $y$ are represented alongside each image.}
\end{figure*}

Equation \ref{eq:2} and thus \ref{eq:4} is valid under the assumption that the overlap of the clouds of both species is 1, which is validated from the CMOS data (Figure \ref{fig:density}). The density $n_j^{(2)}$ of atoms $j$ in the double-species MOT, which can be evaluated with the CMOS camera and the number of atoms.

The calculation of the 2-body interspecies trap-loss collision coefficients is presented in Table \ref{tab:table1}. The calculations are based on the data from the double-species MOT sequences from Section \ref{sec:Double-species MOT}. The main source of uncertainty comes from the estimation of the density, which relies on an extrapolation of the third radius $w_z$, taken as the average between $w_x$ and $w_y$, and the evaluation of the number of atoms in the MOT, which is accurate up to a factor of $\sim 2$. The relative uncertainty on the $\beta_{i,j}$ coefficients is then about $\pm 2\beta_{i,j}$. The 2-body interspecies loss rate coefficients that we obtain are at a similar order of magnitude as the literature \cite{ji_investigation_2011, Holmes20048587Cs}; as \cite{Holmes20048587Cs} typically reports $\beta_{87,Cs}$ of the order of $\sim 1-4 \times10^{-11}\mathrm{cm}^3/\mathrm{s}$ and  $\beta_{85,Cs} \sim 0.5-1.5 \times10^{-11}\mathrm{cm}^3/\mathrm{s}$ in similar experimental conditions. However, our coefficients are about 10 times smaller compared to other work \cite{mancini_trap_2004,harris_magnetic_2008, beta8587}. This disagreement is attributed to different experimental conditions used in these experiments, as \cite{beta8587} reports $\beta_{85,87} \sim 1 \times10^{-11}\mathrm{cm}^3/\mathrm{s}$ for $s = 5$. This dependence on the laser power can be explained by the different collisional processes involved in the interspecies trap-loss coefficients.

\begin{table}[h]
\caption{\label{tab:table1}%
Experimental measurement of the 2-body interspecies loss rate coefficient. The uncertainty on the coefficients is estimated to be a factor of 2-3. Coefficients for ${}^{87}\mathrm{Rb}$/${}^{85}\mathrm{Rb}$ and ${}^{85}\mathrm{Rb}$/${}^{87}\mathrm{Rb}$ are a lower bound estimated from noise on the detection evaluated at $1\times10^6$ atoms. The coefficients are calculated in the double-species MOT.}
\begin{ruledtabular}
\begin{tabular}{lccr}
${}$&
\textrm{Species $i,j$}&
\textrm{$\beta_{i,j}$ [$\mathrm{cm}^3/\mathrm{s}$]}&
${}$\\
\colrule \\
${}$&${}^{87}\mathrm{Rb}$,Cs & $1.1\times10^{-11}$&${}$\\
${}$&Cs,${}^{87}\mathrm{Rb}$ & $1.1\times10^{-11}$&${}$\\
${}$&${}^{85}\mathrm{Rb}$,Cs & $3.2\times10^{-12}$&${}$\\
${}$&Cs,${}^{85}\mathrm{Rb}$ & $2.0\times10^{-12}$&${}$\\
${}$&${}^{85}\mathrm{Rb}$,${}^{87}\mathrm{Rb}$ & $<1.9\times10^{-12}$&${}$\\
${}$&${}^{87}\mathrm{Rb}$,${}^{85}\mathrm{Rb}$ & $<1.1\times10^{-12}$&${}$\\

\end{tabular}
\end{ruledtabular}
\end{table}

The nature of collisions in the multi-species MOT is diverse \cite{weiner_experiments_1999} and highly dependent on the experimental conditions, such as the MOT trap depth. Among the collision types, we denote radiative escape (RE) \cite{radiativeescape}, fine structure-changing (FSC) collisions, hyperfine structure-changing (HFSC) collisions or even photo-association \cite{mancini_trap_2004, telles_trap_2001}. Our measurement shows $\beta_{87,Cs} \approx \beta_{Cs,87}$, as well as $\beta_{85,Cs} \approx \beta_{Cs,85}$, which is not the case in the literature. The fact that $\beta_{Rb,Cs} \neq \beta_{Rb,Cs}$, Rb being  ${}^{85}$Rb or ${}^{87}$Rb, is interpreted as a signature of RE \cite{mancini_trap_2004}. The absence of RE in our experiments could then explain the comparatively lower $\beta_{i,j}$ coefficient that we have measured compared to the literature. We would require more data to properly discriminate between the contributions of HFSC and FSC collisions. Indeed, the behaviour of beta as a function of the MOT intensity gives hints on whether FSC or HFC dominates \cite{santos_hyperfine-changing_1999}. Nonetheless, given our relatively low $s = 1.3$, it is likely that hyperfine structure-changing collision is the dominant process, as HFSC are known to dominate at low intensity \cite{hyperfine_col_s}. HFSC are also found to be dominant in \cite{Holmes20048587Cs} for similar experimental conditions. At low saturation $s$, most of the atoms in the MOT are in their ground state. In such shallow traps, the energy of the hyperfine splitting of the ground state is enough to cause trap loss. The fact that $\beta_{85,Cs}<\beta_{87,Cs}$ in Table \ref{tab:table1} could hereby be explained by the higher hyperfine splitting of $5^{2}\mathrm{S}_{1/2}$ in ${}^{87}\mathrm{Rb}$ than in ${}^{85}\mathrm{Rb}$.

It should be noted that if we use the variations in atom number from the triple-species MOT (Figure \ref{fig:triple}), assuming that they arise from 2-body collisions, we retrieve the same order of magnitude, with no significant difference between $\beta_{i,j}$ and $\beta_{j,i}$. We get $\beta_{85,Cs} = 3.76 \times 10^{-12}$ $\mathrm{cm}^3/\mathrm{s}$ and $\beta_{87,Cs} = 6.37 \times 10^{-12}$ $\mathrm{cm}^3/\mathrm{s}$, taking the average between $\beta_{i,j}$ and $\beta_{j,i}$.
In addition to the 2-body trap-loss coefficients, we also calculated an upper bound on the 3-body trap-loss coefficient $\beta_{i,j,k}$. It adds another loss term in Equation \ref{eq:2} that can be written as $\beta_{i,j,k}n_j^{(3)}n_k^{(3)}N_i$. The expression of $\beta_{i,j,k}$ can be written in a similar shape as Equation \ref{eq:4}:
\begin{eqnarray} \label{eq:5}
\beta_{i,j, k}=\frac{1}{{n_j^{(3)}}n_k^{(3)}\tau_i^{(2)}}\left(\frac{{N_i^{(2)}}^\infty}{{N_i^{(3)}}^\infty}-1\right)
\end{eqnarray}
No change in the number of atoms was observed when switching from a 3-species MOT to a 2-species MOT for any of the three species. The upper bound of $\beta_{i,j,k}$ is then evaluated by considering the noise level on the atom number of $10^{6}$ atoms, and $\tau = 200$ ms. Equation \ref{eq:5} then yields $\beta_{i,j,k} < 1.5\times10^{-23} \mathrm{cm}^{6}.\mathrm{s}^{-1}$, which is typically 2-3 orders of magnitude higher than what is typically observed in much denser systems \cite{3-body-mol, 3-body-impurity}. The current regime does not allow us to observe 3-body collisions.

\section{\label{sec:conclusion}Conclusion and perspectives}

We demonstrated the simultaneous trapping and cooling of ${}^{85}$Rb, ${}^{87}$Rb and ${}^{133}$Cs within the experimental conditions for future cold-atom interferometry with an all-fibre compact and robust laser system. We were able to generate MOTs of about $10^8$ atoms for each species at the saturation intensity $s=1.3$. We investigated the interspecies trap-loss collisions and observed losses between caesium and rubidium. We report averaged interspecies loss-rate coefficients of $\beta_{Cs,87} \approx \beta_{87,Cs}=1.1\times10^{-11}\mathrm{cm}^3/\mathrm{s}$ and $\beta_{85,Cs} \approx \beta_{Cs,85}=2.6\times10^{-12}\mathrm{cm}^3/\mathrm{s}$. No losses between ${}^{85}$Rb and ${}^{87}$Rb were observed, hence we deduced an upper bound of $\beta_{85,87} < 1.9\times10^{-12}\mathrm{cm}^3/\mathrm{s}$. No 3-body trap-loss collisions were observed. The values reported match with the literature, which validates the operation of the original laser system for standard atom cooling and trapping.

This study confirms that our experimental set-up is compatible with future multi-species interferometry experiments. The heteronuclear trap-loss collisions that we measured result in a loss of at most $\approx 7 \%$ of the atoms.

In the considered regime, such inelastic collisions that lead to atom loss are not expected to have a significant impact on the cold-atom interferometer. On the other hand, collisions that do not induce trap loss, such as heteronuclear elastic collisions or $m_F$-changing collisions which were not studied in this paper, can induce collisional frequency shifts and worsen the accuracy of the interferometer. The frequency shift depends on the density of the atomic cloud, which varies through the interferometer because of the thermal expansion of the cloud. As a result, interaction between the atoms could create an additional phase difference in a Mach-Zehnder interferometer \cite{peters_high-precision_2001}. The frequency shift caused by homonuclear collisions in cold-atom interferometers has been extensively studied \cite{87-collision-shift, Cs-collision-shift}, but is still unresolved for heteronuclear collisions.

With the perspective of a simultaneous multi-species interferometer, we expect that the interaction between the three species could impact the performances of the triple-species cold-atom interferometer.

\begin{acknowledgments}
We acknowledge the support of PALM Labex and DSO National Laboratories for funding part of this research activity and the PhD of M. Landru. We thank Goulven Quéméner for fruitful discussion. The authors especially acknowledge C. Diboune and N. Marquet for the early contribution to the experimental setup.
\end{acknowledgments}

\bibliography{apssamp}

\end{document}